\documentclass[german,english,12pt,twoside]{article}
\usepackage{babel}
\usepackage[latin2]{inputenc}
\usepackage{graphicx}

\input CoKon.sty
\begin{document}

\CoKonChapBeg 

\CoKonChapterTitle{Conventional and new directions in studying Cepheids}  

\CoKonChapterAuthor{L\'aszl\'o Szabados}

\CoKonChapterAuthorAffil{Konkoly Observatory of the Hungarian Academy of Sciences\\
P.O. Box 67, H-1525 Budapest, Hungary}

\CoKonAbstract{In the first part of this paper, traditional methods of studying Cepheids are summarized, mentioning Detre's contribution to this field. Then the new directions of Cepheid related research are reviewed with an emphasis on the problems concerning the period-luminosity relationship.}

\CoKonSection{Introduction}

Cepheids are supergiant stars that perform radial pulsation when they cross the classical instability strip in the Hertzsprung-Russell diagram during their post-main sequence evolution. It is the outer layers of Cepheids which oscillate, and the pulsation is maintained by the opacity changes in the partially ionised zones of neutral hydrogen and singly ionised helium capable for transforming heat into mechanical energy via the $\kappa$-mechanism ($\kappa$ is the conventional sign of the opacity).

The mostly monoperiodic Cepheids may seem to be boring targets with respect to the multiperiodic radial and/or non-radial pulsating stars of different types but, in fact, Cepheids are neither perfectly regular, nor homogeneous. Subtle deviations from regularity and homogeneity result in important astronomical consequences. No wonder, Cepheids have remained in the forefront of the variable star studies in spite of the fact that a number of astrophysically important new types of variable stars emerged in the last decades. Due to variety and richness of the relevant studies, selected results are only mentioned in this review.

\CoKonSection{Traditional studies and Detre's contribution}

The Cepheid pulsation is a free oscillation of the star whose frequency corresponds to the eigenfrequency of the stellar plasma sphere. The value of the frequency and its reciprocal, the pulsation period, is governed by the structure of the star, especially by the average density. This dependence gives rise to the well known period-luminosity relationship that exists for various types of pulsating stars. The period-luminosity (P-L) relationship was discovered a century ago by studying Cepheids in the Magellanic Clouds, and since that time this relationship has been instrumental in establishing the extragalactic distance scale.

By the time of mid-20th century, it became obvious that Cepheids are found in at least two varieties: the classical Cepheids belonging to Population~I, and the older Type~II Cepheids which are less luminous at a given pulsation period. In what follows, the term Cepheid covers classical Cepheids only.

The simplest observational study of Cepheids (and any other periodic variable stars) is to obtain photometric data from which the light curve can be constructed and long-term stability of both the light curve and the pulsation period can be investigated. L\'aszl\'o Detre also observed Cepheids -- at the beginning of his career the Cepheids and the RR~Lyrae type variables were not even strictly separated: the RR~Lyrae type stars used to be referred to as short period Cepheids. Detre published three papers on classical Cepheids based on his visual photometric data. The targets of these studies were XY~Cas (Dunst, 1932), SZ~Cas (Dunst, 1933), and YZ~Aur (Detre, 1935).

Two decades later, when the first photoelectric photometer was installed at the Konkoly Observatory, Detre included some bright Cepheids (FF~Aql, SU~Cyg, T~Vul, and U~Vul) in the observational programme. These photoelectric data were only published after Detre's death (Szabados, 1977, 1980).

Long-term changes in the pulsation period can be studied with the help of the  O$-$C method. In the case of Cepheids, the amount and sign of the period variations can serve as an observational check of stellar evolutionary models. During the Cepheid stage of stellar evolution, the supergiant star crosses the instability strip in either direction. A monotonously increasing pulsation period reflects redward crossing of the instability strip, while the blueward transition results in continuously decreasing period. In reality, period fluctuations are often superimposed on the period changes of evolutionary origin. The erratic changes in the pulsation period reflect the physical conditions in the upper layers of Cepheids because this radial pulsation is an atmospheric phenomenon. Detre pioneered the study of period changes of regularly pulsating variable stars. His paper involving O$-$C diagrams for a number of Cepheids was published in 1970. 

\CoKonSection{Role of Cepheids in astronomy}

Importance of Cepheids in astronomy is twofold. In astrophysics they are {\it test objects for stellar evolution theory} (as mentioned above) and models of stellar structure. In extragalactic astronomy and cosmology, Cepheids are considered as {\it primary distance indicators} and the cosmic distance scale is chiefly based on Cepheids. They serve as standard candles because a close correlation exists between the pulsation period and stellar luminosity.  The most recent review on the {\it period-luminosity relationship} is written by Sandage \& Tammann (2006).

In addition to the P-L relationship, Cepheids obey a number of other relationships owing to regularity of their pulsation and some well-known physical principles (e.g. Stefan-Boltzmann law). Close relationships are valid between the period and the spectrum, the period and the radius, the period and the age of the Cepheid, etc. 

Studies of Cepheids have been motivated mostly by the intention of improving any of these relationships either observationally or by theoretical calculations. Nevertheless, existence of period variations is already a hint that one cannot expect perfect regularity in the pulsation of Cepheids. Knowledge of the means how the individual Cepheids deviate from the regular behaviour and the physical explanation of these deviations is essential for the precise calibration of various Cepheid-related relationships, as well as from the point of view of astrophysics.

\CoKonSection{Recent developments}

In spite of the fact that a number of methods suitable for extragalactic distance determination were devised in the last decades, the role of Cepheids as primary distance calibrators has not lessened. The interest in studying Cepheids has been facilitated by the rapid progress in both observational and theoretical astrophysics, especially by the following facts: availability of imaging detectors in the near infrared, existence of massive photometries (with the primary aim at detecting microlensing events), and the enormous increase in computing power.

The advantages of {\it observing Cepheids in infrared} are as follows:\\
- The photometric amplitude at near IR wavelengths is smaller than in the optical spectral region, therefore the mean brightness can be determined reliably from a few observations obtained at random phases. (In the optical band, at least 15-20 observational points are necessary for covering the light curve.) Thus projects aimed at determining extragalactic distances based on the P-L relationship of Cepheids are more feasible in the near IR.\\
- Infrared magnitudes are less affected by interstellar absorption. Thus the P-L relationship has a smaller scatter around the ridge line fit in the near IR photometric bands.\\
- The finite width of the P-L relationship at a given period is partly due to the temperature sensitivity of the pulsation period because the P-L relation is, in fact, period-luminosity-colour (P-L-C) relationship. At longer wavelengths, the monochromatic flux becomes less sensitive to temperature, thus the width of the P-L relation is reduced in the IR bands.\\
- Many Cepheids belong to binary systems whose secondary star is usually a less massive blue or yellow star on or near the main sequence. The photometric contribution from the companion is negligible in the near IR.\\
- The effects of metallicity are much reduced in the infrared as compared with the optical spectral region where metallic absorption lines dominate. Therefore, the Baade-Wesselink method of radius determination can be reliably applied in the near IR. 

The main benefit from the {\it massive photometries}, especially MACHO, OGLE, and EROS, on Cepheid research is the availability of large body of homogeneous photometric data on Cepheids in both Magellanic Clouds. Using these databases, the period dependent behaviour of Cepheids can be studied much more precisely than before. In addition, many new Cepheids were discovered in both satellite galaxies including dozens of double-mode Cepheids (Alcock et~al., 1995; Udalski et~al., 1999a; Soszy\'nski et~al., 2000) and Cepheids which are members in eclipsing binary systems (Udalski et~al., 1999c; Alcock et~al., 2002). The Magellanic double-mode Cepheids outnumber their Galactic counterparts. Their simultaneously excited two modes can be either the fundamental mode and the first overtone, or the first and second overtones. Surprisingly, single-mode Cepheids pulsating in the second overtone have been also found in the Small Magellanic Cloud (Udalski et~al., 1999b). Such stars are not known in our Galaxy, possibly because the mode identification for the Milky Way Cepheids is very difficult and uncertain. 

The increasing number of large telescopes and very sophisticated auxiliary equipments facilitate the discovery of Cepheids in remote galaxies as well as deeper studies of individual Galactic Cepheids. In September 2006 Cepheids were {\it known in 76 galaxies}. Over 5000 Cepheids have been detected beyond the Magellanic Clouds, and about 400 such variables are known in various galaxies of the Virgo Cluster. The number of the known classical Cepheids in our Galaxy only amounts to about 6 per cent of the total known Cepheid sample.

The high precision photometric and deep spectroscopic studies of Galactic Cepheids facilitated a reliable determination of physical properties and surface chemical composition of a large number of these variables. With a spectacular progress in the last decade, {\it abundance determination} based on high resolution spectra has been already performed for more than 150 Cepheids (Kovtyukh et~al., 2005 and references therein). The classical Baade-Wesselink method of {\it radius determination} has been replaced by the infrared surface brightness method (Welch, 1994). The estimated precision of radius determination of Cepheids was about 7 per cent several years ago (Gieren et al., 1999). 

The long-lasting discrepancy between {\it Cepheid masses} derived by various methods was already resolved in early 1990es by using modified opacity values characteristic of stellar interior (Rogers \& Iglesias, 1992; Seaton et~al., 1994). Nevertheless, the agreement is not satisfactory yet. Now, the masses derived from stellar pulsation are smaller by about 10-20 per cent than the mass values deduced from evolutionary models. This problem can be resolved either by more appropriate stellar models or assuming a significant mass loss in pre-Cepheid evolutionary phases (Bono et~al., 2006).

Quite recently, important observational results were achieved on the circumstellar environment of Cepheids: extended envelopes have been found around the brightest Cepheids from near-IR interferometric observations (Kervella et~al., 2006; M\'erand et~al., 2006) testifying that {\it mass loss} occurred in the recent past.

The ample and precise observational data on Cepheids can be used even for studies of {\it star formation history} in particular stellar regions. The period distribution coupled with the period-age relationship is indicative of star formation in the recent past. For example, the period distribution of Cepheids in the Large Magellanic Cloud shows that star formation propagated along the bar of our largest satellite galaxy with a velocity of 100 km\,s$^{-1}$ from SE to NW during the last hundred million years (Alcock et~al., 1999).

Similarly, Cepheids are instrumental in following spatial motion of the point of intersection of two spiral arms in M31. The interaction point could be followed from the age, i.e. the period distribution of Cepheids (Magnier et~al., 1997). The location of intersection now coincides with the superassociation NGC\,206.

Moreover, some characteristics of star formation history in our own galaxy can be investigated, as well. The distribution of metallicity and its gradient as a function of galactocentric radius is an important feature determined from Cepheid studies (Kovtyukh et~al., 2005).

In addition to investigations based on large numbers of Cepheids, {\it studies of individual Cepheid variables} have also resulted in spectacular results. There are quite a few Cepheids that exhibit peculiar behaviour. Polaris ($\alpha$~UMi), the brightest Cepheid, showed a secularly decreasing pulsation amplitude throughout most of the 20th century, but instead of ceasing pulsation, now it oscillates with an extremely small amplitude (Turner et~al., 2005 and references therein). Strangely enough, the locus of Polaris in the H-R diagram is in the middle of the instability strip, so this Cepheid is not about leaving the instability region (Evans et~al., 2002). Another example for secularly declining pulsation amplitude may be the case of Y~Oph (Fernie et~al., 1995). An even more strangely behaving Cepheid is V19 in M33 (Macri et~al., 2001). The tremendous decrease in its pulsation amplitude was accompanied with increasing mean brightness during the 20th century.

In addition to these secular changes, important {\it periodic 
phenomena} also appear among Cepheids. V473~Lyrae, a very short period classical Cepheid (with a pulsation period of 1.491 days), cyclically varies its amplitude by a factor of about 15. The modulation period is as long as 1200 days (Burki et~al., 1986 and references therein). The physical cause of this unprecedented behaviour has not been clarified yet.

Another type of independent periodicity is due to the orbital motion, if the Cepheid is a member in a binary system. The frequency of occurrence of {\it binaries among Cepheids} is as large as incidence of binarity among common stars in the solar neighbourhood, i.e. {\it exceeds 50 per cent} (Szabados, 2003). Such a high percentage was not foreseen earlier. Cepheids belonging to binary systems are key objects for determining the physical properties of Cepheid variables, including stellar luminosity which facilitates reliable calibration of the zero-point of the P-L relationship (e.g. Evans, 1992). Especially valuable are in this respect the eclipsing systems involving a Cepheid component, because the inclination of the orbital plane follows from the eclipsing nature, and knowledge of the inclination removes uncertainty in the mass determination. Unfortunately, such pairs have not been found in our Galaxy but three {\it eclipsing binary systems with Cepheid primaries} have been revealed in the LMC (Alcock et~al., 2002; Udalski et~al., 1999b). 

The {\it pairs consisting of two Cepheid variables} are extremely interesting objects from the viewpoint of stellar evolution. In addition to the archetype, CE~Cas, Alcock et~al. (1995) revealed three such pairs in the LMC, while Udalski et~al. (1999a) detected one pair in the SMC. Though the two components cannot be separated, it is clear from the period ratio of the two excited oscillations that the observed variations cannot be explained with double-mode pulsation of a solitary Cepheid in any of these cases.

In the {\it beat Cepheids}, the two excited oscillations are not independent of each other: they correspond to low order radial modes of stellar pulsation. Though a number of faint Galactic double-mode Cepheids have been discovered in the last decade, the number of such variables is only slightly over twenty, i.e. much less than their known counterparts discovered from the data collected during the photometries of either Magellanic Cloud. Quite recently, double-mode Cepheids were discovered in M33 (Beaulieu et~al., 2006) among the huge sample of the photometric survey performed by Hartman et~al. (2006). The survey covering the whole area of this more remote galaxy resulted in identifying about 2000 Cepheids in M33 (Hartman et~al., 2006).
There has been a spectacular {\it progress in modelling} double-mode pulsation in Cepheids, too. While purely radiative models have failed to reproduce simultaneous double-mode periodicity of Cepheids for decades, when taking into account turbulent convection in the hydrodynamic calculations, Koll\'ath et~al. (2002) succeeded in obtaining a stable beat Cepheid behaviour.

Another major result among the Cepheid related theoretical investigations is the confirmation of existence of strange Cepheids. Stars performing surface mode pulsation were predicted by Buchler et~al. (1997), and the first representatives of such short period, ultralow amplitude variables were discovered from the MACHO photometry of the LMC by Buchler et~al. (2005). Discovery of non-radial oscillations as well as triple-mode pulsation in classical Cepheids were also announced based on the OGLE LMC data (Moskalik et~al., 2004).

\CoKonSection{Problem of universality of the P-L relationship}

There are quite a few effects that place the individual Cepheids scattered around the ridge line P-L(-C) relationship, thus resulting in a finite width  of this plot. The most important effects being:\\
- interstellar reddening and absorption;\\
- presence of a companion star;\\
- mass loss;\\
- magnetic field;\\
- mode of pulsation;\\
- nonlinearity of the relation;\\
- differences in the chemical composition.

The amount and effects of interstellar absorption has been widely discussed and thoroughly studied in the Cepheid related literature. For Galactic Cepheids, the reddening correction and the intrinsic colour index is determined individually. In the case of extragalactic Cepheids, however, this is not a viable procedure, and, instead, the practically reddening-free Wesenheit function, $W$, is used (see Madore, 1976):

$W = \langle V \rangle - R (\langle B \rangle - \langle V \rangle)$

\noindent where $R$ is the ratio of total-to-selective absorption. The assumption that $R$ is constant throughout any galaxy is only a rough approximation. For extragalactic Cepheids, the absorption consists of two parts: internal absorption in the galaxy hosting the Cepheid and foreground absorption produced by interstellar matter along the given line of sight in our Galaxy. The effect of this latter component can be readily determined by multicolour photometry (Freedman \& Madore, 1990 and references therein).

The photometric effect of possible companion stars (either physical or optical companions) is usually not taken into account. Neglect of binarity may lead to a systematic error in determining the luminosity of Cepheids useful for the calibration of the P-L relationship (Szabados, 1997), while line-of sight companions in crowded stellar fields of remote galaxies falsify the distance modulus derived for the given system.

Studies on the mass loss can gather a new impetus by recent discoveries of envelopes around bright Cepheids (Kervella et~al., 2006; M\'erand et~al., 2006). 

Existence of magnetic field and its effect on luminosity of Cepheids is a topic of worthy of closer attention from both theoretical and observational points of view.

The pulsation mode of extragalactic Cepheids can be determined relatively simply because stars oscillating in different modes are situated along distinct P-L relationships. In the case of Galactic Cepheids, however, the determination of the pulsation mode is not easy. There are contradictory propositions on the pulsation mode of some well studied bright Cepheids.
Difficulties in the mode identification may also cause that no singly periodic Cepheid pulsating in the second overtone is known in our Galaxy, while a plenty of such stars have been found in the SMC (Udalski et~al., 1999b). 

Quite recently, it turned out that the P-L relationship of the LMC is nonlinear, showing a break at the pulsation period of about ten days (Sandage et~al., 2004; Ngeow \& Kanbur, 2006 and references therein). This effect is caused by nonlinearity of the period-colour relation and has its physical origin in the interaction of the hydrogen ionization front with the Cepheid photosphere. This interaction changes with the phase of pulsation and metallicity producing the observed changes in the Cepheid P-C and P-L relationships. Note that nonlinearity is characteristic of the relationships of the metal poor LMC, while the corresponding relations valid for the Milky Way galaxy are linear.

The role of {\it metallicity} in modifying the relationships valid for Cepheids is the key issue in the recent Cepheid related literature. The very precisely measurable period ratio of double-mode Cepheids clearly depends on the abundance of the heavy elements as shown by the comparison of Galactic beat Cepheids and their siblings in the Large Magellanic Cloud (Alcock et~al., 1999). The [Fe/H] values of Galactic beat Cepheids determined individually from high resolution spectra also confirm existence of metallicity dependence of the period ratio (Szil\'adi et~al., 2006). Moreover, Klagyivik \& Szabados (2006) pointed out that some phenomenological properties of Cepheids, e.g. ratio of amplitudes of photometric and radial velocity variations also depend on the heavy element abundance, $Z$.

Nevertheless, the most important problem in this respect is the dependence of the zero point and the slope of the P-L relationship on metallicity. The era of contradictory results has not been over yet. A numerical parameter, $\gamma$, describing this metallicity dependence has been introduced by Sakai et al. (2004): $\gamma = {\delta}(m-M)/{\delta}\log Z$ where
${\delta}(m-M) = (m-M)_Z -(m-M)_0$
is the difference of distance modulus corrected for the effect of metallicity and the uncorrected value, and
$ {\delta}\log Z = (\log Z)_{LMC} - (\log Z)_{\rm extragal}$.
The most recent studies (Sakai et~al., 2004) resulted in
$\gamma = -0.24 \pm 0.05$ mag/dex. It is worthy to mention that  Freedman et al. (2001) used practically the same value of the $\gamma$ in the final paper on the HST Key Project on the Hubble constant. However, theoretical models calculated by Romaniello et~al. (2005), taking into account the variable He content, are not compatible with these observational findings. 

In order to determine a reliable value of the distance modulus of the galaxy, at least the average metallicity of the host galaxy has to be known, in any case. Caputo et~al.'s (2004) new method is a promising development in this respect. They pointed out that the luminosity difference between the RR~Lyrae type variables and the more massive pulsators with the same period is a function of metallicity. The more massive, short period pulsators involved in this method are the so called {\it anomalous Cepheids}. According to the new paradigm, however, the anomalous Cepheids are classical Cepheids with extremely low metal content (Caputo et~al., 2004; Marconi et~al., 2004).

Knowledge of metallicity of Cepheids is, therefore, especially important for the precise calibration of the P-L relationship, i.e. to fix the bottom rung of the cosmic distance ladder.

\CoKonAck Cepheid related studies at the Konkoly Observatory are partly supported by the Hungarian OTKA grant T046207.

\CoKonReferences

Alcock C., Allsman R. A., Axelrod T. S., et al. (The MACHO Collaboration), 1995, {\it AJ}, {\bf 109}, 1653

Alcock C., Allsman R. A., Alves D. R., et al. (The MACHO Collaboration), 1999, {\it AJ}, {\bf 117}, 920 

Alcock C., Allsman R. A., Alves D. R., et al. (The MACHO Collaboration), 2002, {\it ApJ}, {\bf 573}, 338

Beaulieu J.-P., Buchler J.-R., Marquette J.-B., et al., 2006, {\it ApJ}, (accepted), \hfill\break astro-ph/0610749

Bono G., Caputo F., \& Castellani V., 2006, {\it MemSAIt}, {\bf 77}, 207

Buchler J. R., Wood R. R., Keller S., \& Soszy\'nski I., 2005, {\it ApJ}, {\bf 631}, L151

Buchler J. R., Yecko P. A., \& Koll\'ath Z., 1997, {\it A\&A}, {\bf 326}, 669

Burki G., Schmidt E. G., Arellano Ferro A., et al., 1986, {\it A\&A}, {\bf 168}, 139

Caputo F., Castellani V., Degl'Innocenti S., Fiorentino G., \& Marconi M., 2004, {\it A\&A}, {\bf 424}, 927

Dunst L., 1932, {\it Astr. Nachr.}, {\bf 246}, 361

Dunst L., 1933, {\it Astr. Nachr.}, {\bf 247}, 309

Detre L., 1935, {\it Astr. Nachr.}, {\bf 257}, 361

Detre L., 1970, {\it Ann. Univ-Sternw. Wien}, {\bf 29}, No.\,2, 79

Evans N. R., 1992, {\it ApJ}, {\bf 389}, 657

Evans N. R., Sasselov D. D., \& Short C. I., 2002, {\it ApJ}, {\bf 567}, 1121

Fernie J. D., Khosnevissan M. H., \& Seager S., 1995, {\it AJ}, {\bf 110}, 1326

Freedman W. L. \& Madore B. F., 1990, {\it ApJ}, {\bf 365}, 186

Freedman W. L., Madore B. F., Gibson B. K. et~al., 2001, {\it ApJ}, {\bf 553}, 47

Gieren W. P., Moffett T. J., \& Barnes T. G. III, 1999, {\it ApJ}, {\bf 512}, 553

Hartman J. D., Bersier D., Stanek K. Z., et~al., 2006, {\it MNRAS}, {\bf 371}, 1405

Kervella P., M\'erand A., Perrin G., Coud\'e du Foresto V., 2006, {\it A\&A}, {\bf 448}, 623

Klagyivik P. \& Szabados L., 2006, {\it Publ. Astron. Dept. of E\"otv\"os Univ.}, Budapest, {\bf 17}, 121

Koll\'ath Z., Buchler J. R., Szab\'o R., \& Csubry Z., 2002, {\it A\&A}, {\bf 385}, 932

Kovtyukh V. V., Wallerstein G., \& Andrievsky S. M., 2005, {\it PASP}, {\bf 117}, 1173

Macri L. M., Sasselov D. D., \& Stanek K. Z., 2001, {\it ApJ}, {\bf 550}, L159

Madore B. F., 1976, {\it RGO Bull.}, No\,182, 153

Magnier E. A., Prins S., \& Augusteijn T., 1997, {\it A\&A}, {\bf 326}, 442

Marconi M., Fiorentino G., \& Caputo F., 2004, {\it A\&A}, {\bf 417}, 1101

M\'erand A., Kervella P., Coud\'e du Foresto V., et al., 2006, {\it A\&A}, {\bf 453}, 155

Moskalik P., Ko\l{}aczkowski Z., \& Mizerski T., 2004, in Proc. IAU Coll. 193, {\it Variable Stars in the Local Group}, eds. D. W. Kurtz \& K. R. Pollard, ASPC 310 (San Francisco: ASP), 498

Ngeow C. \& Kanbur S. M. 2006, {\it ApJ}, {\bf 650}, 180

Rogers F. J. \& Iglesias C. A., 1992, {\it ApJS}, {\bf 79}, 507

Romaniello M., Primas F., Mottini M., Groenewegen M., Bono G., \& Fran\c{c}ois P., 2005, {\it A\&A}, {\bf 429}, L37

Sakai S., Ferrarese L., Kennicutt R. C. Jr., \& Saha A., 2004, {\it ApJ}, {\bf 608}, 42

Sandage A. \& Tammann G. A., 2006, {\it ARA\&A}, {\bf 44}, 93

Sandage A., Tammann G. A., \& Reindl B., 2004, {\it A\&A}, {\bf 424}, 43

Seaton M. J., Yan Y., Mihalas D., \& Pradhan A. K., 1994, {\it MNRAS}, {\bf 266}, 805

Soszy\'nski I., Udalski A., Szyma\'nski M., et al., 2000, {\it AcA}, {\bf 50}, 451

Szabados L., 1977, {\it Mitt. Sternw. ung. Akad. Wiss.}, Budapest, No.\,70.

Szabados L., 1980, {\it Commun. Konkoly Obs. Hung. Acad. Sci.}, Budapest, No.\,76.

Szabados L., 1997, in Proc. Conf. {\it HIPPARCOS Venice'97}, ed. B. Battrick, ESA SP-402, 657

Szabados L., 2003, {\it IBVS}, No.\,5394

Szil\'adi K., Vink\'o J., Poretti E., et al., 2006, in preparation

Turner D. G., Savoy J., Derrah J., et al., 2005, {\it PASP}, {\bf 117}, 207

Udalski A., Soszy\'nski I., Szyma\'nski M., et al., 1999a, {\it AcA}, {\bf 49}, 1

Udalski A., Soszy\'nski I., Szyma\'nski M., et al., 1999b, {\it AcA}, {\bf 49}, 45

Udalski A., Soszy\'nski I., Szyma\'nski M., et al., 1999c, {\it AcA}, {\bf 49}, 223

Welch D. L., 1994, {\it AJ}, {\bf 108}, 1421

\CoKonEndreferences

\CoKonChapEnd

\end{document}